\documentclass[aps,reprint,twocolumn,groupedaddress]{revtex4-2}

\usepackage{amsmath, amssymb, natbib, graphicx, url, cleveref}
\Crefformat{figure}{#2Fig.~#1#3}

\usepackage{xcolor, subfigure}

\newcommand\todo[1]{\textcolor{red}{#1}}

\usepackage{chngcntr}
\counterwithout{paragraph}{subsubsection}

\begin{document}


\title{Kinetic phase diagram for nucleation and growth of competing crystal polymorphs in charged colloidal suspensions}


\author{Willem Gispen}
\email[]{w.h.gispen2@uu.nl}

\author{Marjolein Dijkstra}
\email[]{m.dijkstra@uu.nl}

\affiliation{Soft Condensed Matter, Debye Institute for Nanomaterials Science, Utrecht University}


\date{\today}

\begin{abstract}
We determine the full kinetic phase diagram for nucleation and growth of crystal phases in a suspension of charged colloids. We calculate nucleation barrier heights for face-centered cubic (fcc) and body-centered cubic (bcc) crystal phases for varying screening lengths and supersaturations using the seeding approach in extensive simulations. Using classical nucleation theory, we determine for the entire metastable fluid region the crystal polymorph with the lowest nucleation barrier. Surprisingly, we find a regime close to the triple point where metastable bcc can form due to a lower nucleation barrier, even though fcc is the stable phase. For higher supersaturation, we find that the difference in barrier heights decreases and we observe a mix of hexagonal close-packed (hcp), fcc and bcc structures in the growth of crystalline seeds as well as in spontaneously formed crystals.

\end{abstract}


\maketitle




Crystallization plays a prominent role in many research areas and industrial processes, including weather prediction, protein characterization and pharmaceutical drugs production. However, the kinetic pathways of nucleation and the mechanisms of polymorph selection during crystallization are far from being  well-understood. For example, the end product of crystallization is not necessarily the stable structure, but can even be an undesired phase. More complicated scenarios are also possible with structural transformations taking place at various stages of the crystallization process. 

Colloidal suspensions are ideal for studying nucleation and crystallization  as the particle coordinates can be tracked  by advanced microscopy due to the relatively large size and slow diffusion of the colloids \cite{palberg_crystallization_2014, herlach_overview_2016, wette_nucleation_2007}.
Charge-stabilized colloids are specifically suited for studying the selection between crystal polymorphs, since they show an intriguing competition between face-centered cubic (fcc) and body-centered cubic (bcc) crystal phases. However, the crystallization mechanism in charged colloids is not clear-cut and numerous experimental observations are hitherto unexplained. For instance, the observation of broad fluid-solid and fcc-bcc coexistences  in various experiments  \cite{schope_response_1998, bareigts_packing_2020} is inconsistent with the theoretical phase diagrams \cite{hynninen_phase_2003} that predict  narrow phase coexistences. Furthermore, a wide variety of crystallization mechanisms has been observed in experiments, ranging from a simple one-step nucleation mechanism of fcc crystals \cite{dhont1992time,gasser_real-space_2001}, to the emergence of metastable bcc crystals that subsequently transform into fcc \cite{xu_formation_2010, zhou_kinetics_2011}, as well as the emergence of hcp before fcc is formed  \cite{tan_visualizing_2014}.

Simulations do not seem to reach consensus either, as they report conflicting results such as the observation of  predominantly bcc-structured  (pre)critical  nuclei in regions where fcc is stable \cite{auer_crystallization_2002, blaak_crystal_2004}, a two-stage fluid-fcc crystallization via an intermediate bcc phase \cite{kratzer_two-stage_2015}, formation of bcc-ordered precursors \cite{ji_crystal_2018}, or the formation of  metastable bcc with numerous cross-nucleations of hcp on stable fcc and fcc on metastable hcp crystals  \cite{desgranges2007polymorph}. It is important to note that  simulations of crystallization are prohibitively slow because nucleation is a rare event. Hence, simulation studies on nucleation are limited to only a few state points and interaction parameters, making it difficult to obtain a coherent picture of the different nucleation mechanisms. 
Moreover, these simulations can only be performed at high supersaturations.

To date, we can only rely on simple guidelines to predict how a system crystallizes. In 1879, Ostwald formulated his famous step rule that the phase that nucleates need not be the stable phase, but may also be the phase that is closest in free energy to the metastable fluid phase, i.e. the less stable polymorph.  
This would result in a complete reversal of the thermodynamic phase diagram, i.e. fcc nucleates when bcc is stable and bcc forms when fcc is stable. In the 1930s, Stranski and Totomanov conjectured  that the phase that nucleates should have the lowest free-energy barrier with the fluid phase, which can be different from the stable phase. Finally, Alexander and McTague argued on the basis of Landau theory and general symmetry considerations that nucleation of bcc is always favored at low supersaturations in the case of weakly first-order freezing transitions  \cite{alexander_should_1978}. It is clear that these rules of thumb are too general to be universally valid. For instance, at high screening close to the hard-sphere limit, bcc is mechanically unstable thereby violating Alexander and McTague's conjecture.  


\begin{figure}[!t]
    \centering
    \includegraphics[width=\linewidth]{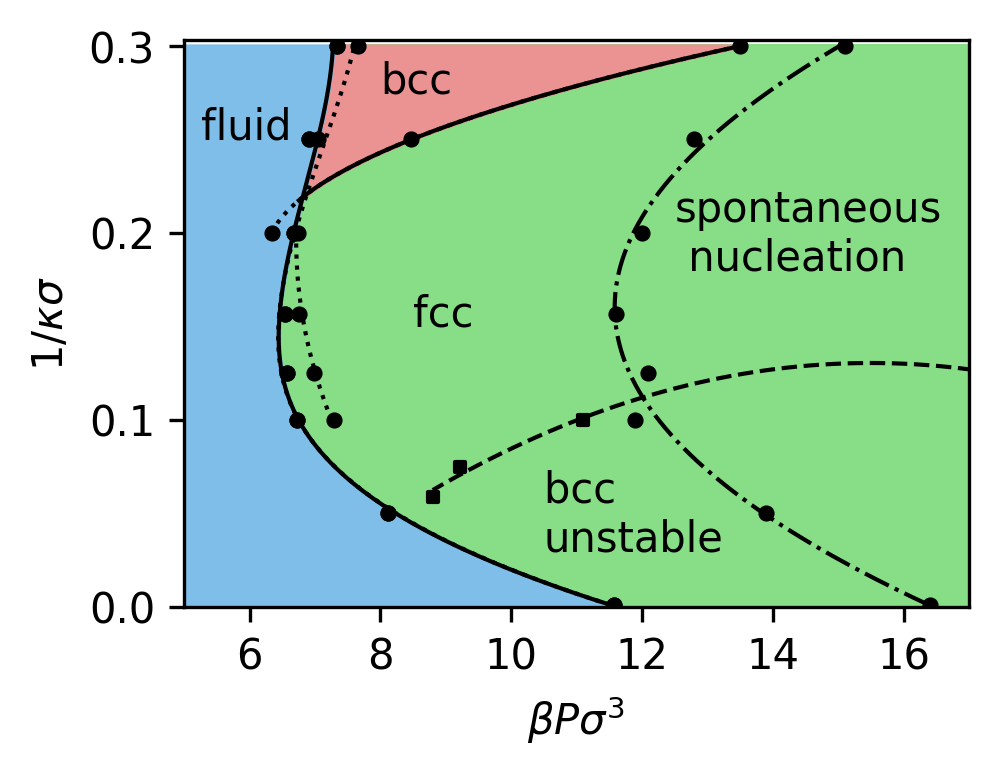}
    \caption{Bulk  phase diagram of highly charged colloids with a  contact value $\beta\epsilon=81$ in the pressure $\beta P \sigma^3$ - Debye screening length $1/\kappa\sigma$ plane. The fluid-fcc, fluid-bcc, and bcc-fcc  binodals are denoted by solid lines and their metastable extensions by  dotted lines. The dash-dotted line marks the pressure at which the fluid spontaneously crystallizes. The dashed line indicates where the bcc phase spontaneously transforms into fcc.  Dots are the actual measurements, lines are spline interpolations to guide the eye.
    \todo{}
    }
    \label{fig:phase-diagram}
\end{figure}

\begin{figure}[!t]
  \includegraphics[width=\linewidth]{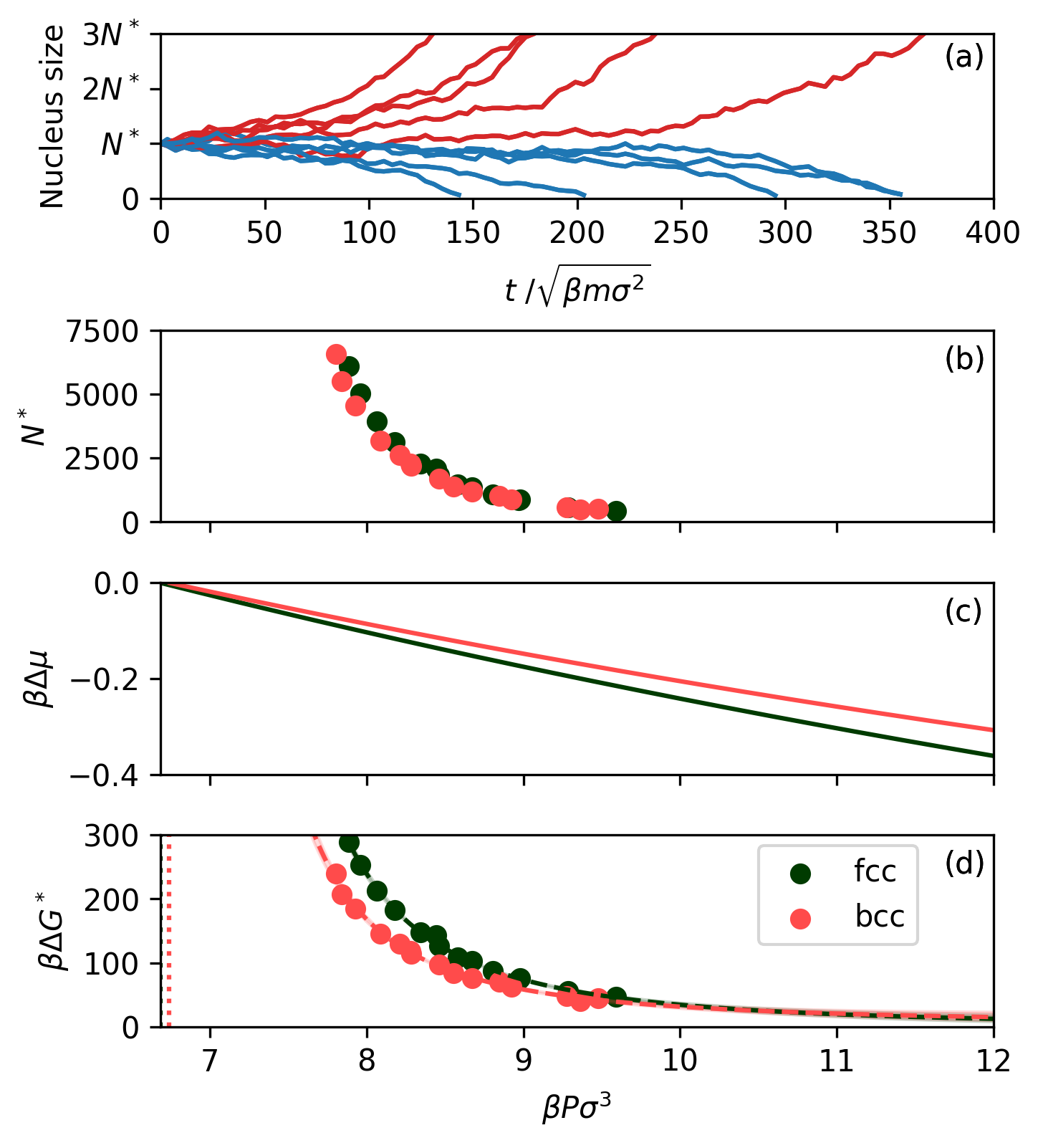}
  \caption{(a) Largest cluster size of a bcc seed with an initial size $N^*=2300$ as a function of time $t/\sqrt{\beta m \sigma^2}$ using the seeding technique in 10 independent simulations of charged colloids at a  screening length $1/\kappa\sigma = 0.2$.  (b) The critical nucleus size $N^*$ of bcc and fcc, (c)  supersaturation $\beta \Delta \mu=\beta (\mu_{x}-\mu_{f})$ with $\mu_{x}$ and  $\mu_{f}$ the chemical potential of the crystal and fluid phase, respectively,  and (d) the Gibbs free-energy barrier  $\beta \Delta G^*$, all as a function of pressure $\beta P\sigma^3$.}
  \label{fig:seeding}
\end{figure}

In this Letter, we present a coherent picture of the different crystallization scenarios of charged colloids. We first determine the equilibrium phase diagram as a function of screening length  and pressure  using free-energy calculations. We then calculate the nucleation barrier heights for the full region of the phase diagram where fcc is the thermodynamically stable phase  using extensive simulations based on the seeding technique \cite{espinosa_seeding_2016}. The seeding approach allows us not only to determine nucleation barriers at relatively low supersaturation but also to compare the barrier heights of competing crystal structures. We then characterize the structure of growing crystals obtained from both seeded and brute-force simulations. 
In this way, we obtain a kinetic phase diagram containing information about  the nucleation as well as the growth stages of crystallization.


We consider a charge-stabilized colloidal suspension, which is well-described by a system where the electrostatic interactions between the  colloids are described by  a screened Coulomb (Yukawa) potential  
\begin{equation*}
    \beta u_Y (r) = \beta \epsilon \frac{\exp\lbrack-\kappa\sigma(r/\sigma - 1)\rbrack}{r/\sigma},
\end{equation*}
with  $\beta\epsilon$ the contact value and  $1/\kappa\sigma$  the Debye screening length  which  determine the strength  and   range of the repulsion, respectively. 
The excluded-volume interactions between the colloids are represented by a pseudo-hard-core potential $\beta u_{PHS} (r)$ which reproduces well the hard-sphere equation of state  \cite{jover_pseudo_2012}. The total  interaction potential of the pseudo-hard-core Yukawa system reads $\beta u(r) = \beta u_Y (r) + \beta u_{PHS} (r)$. We set $\beta\epsilon=81$ throughout this paper. We note  that for this high contact value the phase behavior of this pseudo-hard-core Yukawa system can be mapped onto that of point Yukawa particles as shown in Ref.\ \cite{hynninen_phase_2003}. Consequently, our results are valid for any  contact  value that is sufficiently high, i.e. $\beta \epsilon>20$, by exploiting the mapping of point Yukawa particles onto  hard-core Yukawa particles \cite{hynninen_phase_2003}. %

%



\begin{figure*}[!htbp]
    \centering
    \includegraphics[width=\textwidth]{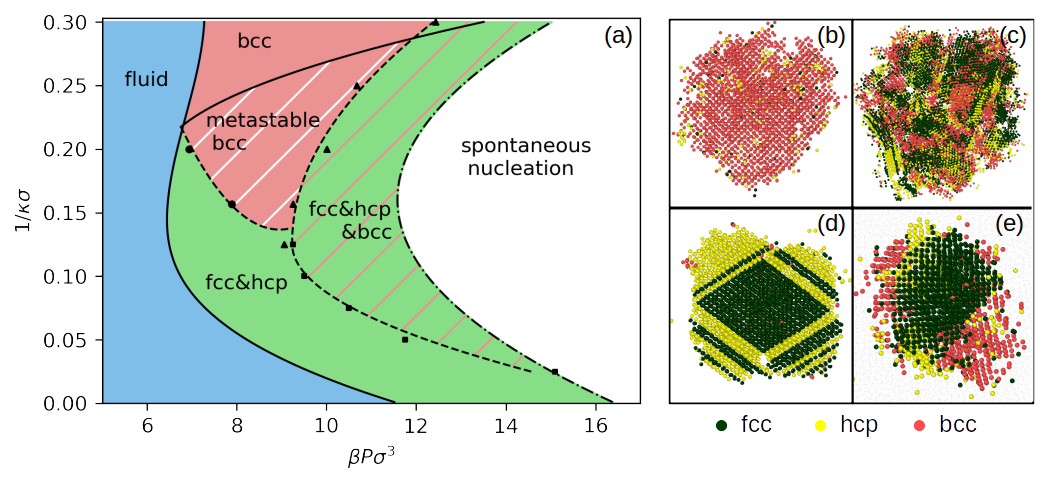}
    \caption{
    (a) Kinetic phase diagram of highly charged colloids  with a contact value $\beta\epsilon=81$ in the pressure $\beta P \sigma^3$ -- Debye screening length $1/\kappa\sigma$ plane. The fluid-fcc, fluid-bcc, and bcc-fcc binodals are denoted by solid lines, and the dash-dotted line marks the pressure beyond which the fluid spontaneously crystallizes.  The red (green) region denotes the region where bcc (fcc) has a lower nucleation barrier than fcc (bcc). The red region with stripes near the triple point denotes the metastable bcc region, where bcc has a lower nucleation barrier while being metastable.  The triangles mark the pressures above which the nucleation barriers of fcc and bcc are indistinguishable within our statistical accuracy, while the squares mark the pressures above which a mix of fcc, hcp and bcc is observed in the growth of crystal seeds. The dashed line is a spline interpolation of the triangles and squares, and therefore indicates the onset of a mixed fcc/hcp/bcc region (green with red stripes). Dots are actual measurements and lines are spline interpolations to guide the eye. (b-e) Cross-sections of crystal nuclei illustrating the different regions in the kinetic phase diagram.  The color coding shown below the snapshots denotes the local structure as recognized by the polyhedral template matching, while  unrecognized particles are reduced in size. (b) A pure bcc crystal grown from a bcc seed at $(\beta P \sigma^3, 1/\kappa\sigma) = (8.2, 0.2)$,  (c) a mix of fcc, hcp, and bcc grains formed spontaneously at $(\beta P \sigma^3, 1/\kappa\sigma) = (11, 0.157)$, (d) a mix of fcc and hcp grown from an fcc seed at $(\beta P \sigma^3, 1/\kappa\sigma) = (8.2, 0.1)$, and (e) a mix of fcc, hcp and bcc grown from an fcc seed at $(\beta P \sigma^3, 1/\kappa\sigma) = (8.6, 0.157)$.}
    \label{fig:metastable-phase-diagram}
\end{figure*}

We determine the bulk equilibrium phase diagram for charged colloids with a contact value $\beta \epsilon=81$ using free-energy calculations, see Supplemental Materials (SM) \cite{SM} for technical details. We present the phase diagram in the reduced pressure $\beta P\sigma^3$ - Debye screening length $1/\kappa \sigma$ representation in Fig. \ref{fig:phase-diagram}. The phase diagram displays  fluid-fcc, fluid-bcc, and bcc-fcc binodals and their metastable extensions, denoted by solid  and dotted lines, respectively, as well as a triple point at Debye screening length $1/\kappa\sigma \approx 0.22$ and pressure $\beta P\sigma^3 \approx 6.7$ in good agreement with Ref.\ \cite{hynninen_phase_2003}. Hence, the bcc phase is only stable for $1/\kappa\sigma \gtrsim 0.22$. In addition, we identify the stability regions of the fluid and the bcc phase, by 
determining at which pressure the fluid spontaneously crystallizes and the bcc spontaneously transforms into fcc. We denote the boundaries where fluid and bcc become unstable by a dash-dotted and dashed line, respectively. It is clear that at high screening the Alexander-McTague conjecture stating that  nucleation of bcc should be favored near  melting is violated  as the bcc phase is simply unstable. Moreover, both the Alexander-McTague conjecture as well as    Ostwald's step rule stating that the least stable polymorph should nucleate first cannot be valid in the region between the fluid-fcc binodal and the metastable fluid-bcc binodal as the bcc phase has a higher Gibbs free energy  than the fluid phase, see also SM \cite{SM}. 

To study the kinetic competition between  fcc and bcc crystal polymorphs, we use a method similar to recent work on metastable phases in iron \cite{sadigh_metastablesolid_2021}. Compared to Ref.\ \cite{sadigh_metastablesolid_2021}, we build our method more explicitly on the seeding technique \cite{espinosa_seeding_2016}, which has been used and validated in many different systems over the past few years, e.g.\ in hard spheres \cite{sanchez-burgos_fcc_2021}, oppositely charged colloids \cite{sanchez-burgos_parasitic_2021}, and NaCl \cite{espinosa_crystal-fluid_2015}.
The seeding technique allows us to efficiently measure the nucleation barriers for relatively low supersaturation, but more importantly to also compare directly the nucleation barriers of fcc and bcc. The method relies on the combination of molecular dynamics simulations with classical nucleation theory (CNT). According to CNT, the Gibbs free-energy barrier height $\Delta G^*$ is related to the supersaturation $|\Delta \mu|=|\mu_{x}-\mu_{f}|$, i.e. the difference in chemical potential between the  stable crystal $\mu_x$ and supersaturated fluid phase $\mu_f$ as 
\begin{equation}
    \label{eq:cnt}
    \Delta G^* = \frac{1}{2} N^* |\Delta \mu|,
\end{equation}
where $N^*$ is the number of particles of the critical nucleus at the top of the Gibbs free-energy barrier.  In the seeding approach, we insert  a seed of the crystal structure of interest, either bcc or fcc, in a metastable fluid phase. After carefully equilibrating the crystal seed and its interface with the fluid, we simulate the system for a range of pressures to determine at which pressure, i.e. the critical pressure $P^*$, the seed will grow or melt with equal probability, while  the crystalline seeds will predominately melt for  $P < P^*$, and grow for $P > P^*$. In   \Cref{fig:seeding}a, we exemplarily show that a bcc seed of size $N^*\simeq 2300$ melts or grows with $50\%$ probability at a  critical pressure $\beta P^* \sigma^3 \approx 8.28$. 
Subsequently, we obtain the nucleation barrier $\Delta G^*$ for this critical nucleus size $N^*$ using Eq. \eqref{eq:cnt} with  $|\Delta \mu|$  the supersaturation at this critical pressure $P^*$.  Using fcc and bcc seeds of many different sizes, the seeding approach enables us to  determine the nucleation barriers $\Delta G^*$ of both crystal polymorphs for supersaturations $|\Delta \mu|$ close to bulk coexistence. Using classical nucleation theory, these nucleation barriers can be fitted and extrapolated to the entire metastable fluid region (see \Cref{fig:seeding}d and SM \cite{SM}). 


We present our seeding simulation results  in a kinetic phase diagram in  \Cref{fig:metastable-phase-diagram}. The kinetic phase diagram shows a region denoted by red where bcc has a lower nucleation barrier than fcc, and a green region, where fcc has a lower nucleation barrier than bcc. For sufficiently high  pressures (marked with triangles)
the nucleation barriers of fcc and bcc become indistinguishable within our statistical accuracy.
Interestingly, there is a region near the triple point where bcc has a lower nucleation barrier than fcc even though it is metastable. This marked result can be explained by a lower interfacial free energy of the fluid with  bcc  compared to that with  fcc, and is thus  a manifestation of Ostwald's step rule and Alexander and McTague's conjecture. On the other hand, in a large region near the fluid-fcc binodal, the stable fcc phase has a lower nucleation barrier than the metastable bcc phase. Therefore, in this region both Ostwald's step rule and Alexander and McTague's conjecture are violated.


Finally, we turn our attention to the crystal growth regime. We select the seeding simulations that resulted in crystal growth, and use only simulations with pressures $P$ close to the critical pressure ($\beta|P-P^*|\sigma^3 < 0.2$).
We determine the structural composition of the resulting crystals using polyhedral template matching \cite{larsen_robust_2016}. For low supersaturations, i.e.\ for pressures close to bulk coexistence, relatively pure crystals are observed. More precisely, we  observe that  seeds with the crystal structure corresponding to  the lowest nucleation barrier retain their initial crystal  structure during  growth. Notably, bcc seeds in the metastable bcc region grow out to pure bcc crystals as shown in \Cref{fig:metastable-phase-diagram}b, demonstrating  that a proper metastable bcc phase forms in this region. For high screening $1/\kappa\sigma < 0.15$ and low supersaturations, fcc seeds grow out into a mixture of fcc and hcp due to stacking faults as shown in  \Cref{fig:metastable-phase-diagram}d. On the other hand, there is a large region at higher supersaturation, where  fcc and bcc seeds grow out into a polycrystalline mixture of fcc, hcp, and bcc grains.   Interestingly, this region corresponds exactly to  the region where the nucleation barriers as determined from the seeding approach become indistinguishable (see SM \cite{SM}).
Additionally, we find that in this region, the fraction of bcc increases with pressure and $1/\kappa\sigma$.

\begin{figure}
    \centering
    \includegraphics[width=\linewidth]{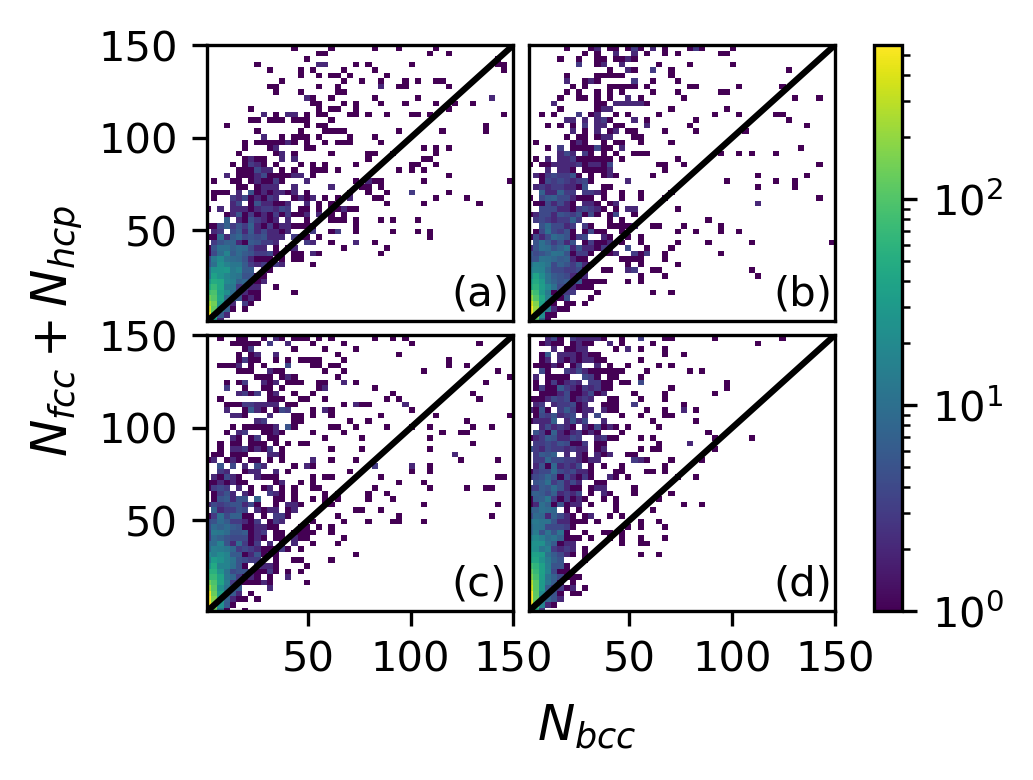}
    \caption{Two-dimensional histogram of the structure of spontaneously formed crystal nuclei as recognized by  polyhedral template matching \cite{larsen_robust_2016} for varying screening lengths and pressures $(1/\kappa\sigma, \beta P \sigma^3)$= (0.3,15.5) (a), (0.2,12.5) (b), (0.157,11.5) (c), and (0.1,12.0) (d) as a function of the number of fcc and hcp particles, $N_\text{fcc} + N_\text{hcp}$, and bcc particles $N_\text{bcc}$. The black diagonal line corresponds to $N_\text{fcc} + N_\text{hcp} = N_\text{bcc}$, indicating the distinction between fcc/hcp and bcc dominated crystal nuclei. } 
    \label{fig:spontaneous-histogram}
\end{figure}

To test our predictions from the seeding simulations, we also perform brute-force crystallization simulations. We  again find in agreement with the seeding simulations that the resulting crystals consist of a mixture of fcc, hcp, and bcc grains as determined by polyhedral template matching. To quantify this further, we determine the composition of the crystals by counting the number of bcc and fcc/hcp particles in the observed nuclei. In \Cref{fig:spontaneous-histogram}, we plot the probability to observe a crystal cluster consisting of $N_{\textrm{bcc}}$ and $N_{\textrm{fcc}}+N_{\textrm{hcp}}$ particles in a  two-dimensional histogram. \Cref{fig:spontaneous-histogram} clearly shows that the structure of spontaneously formed nuclei are dominated by  fcc and hcp  in both the nucleation and growth regime, but there are also bcc grains present, even in very large nuclei. As an example, \Cref{fig:metastable-phase-diagram}c shows a cross-section of a spontaneously formed crystal  at a pressure $\beta P \sigma^3 = 11$ and screening length $1/\kappa \sigma = 0.157$. This nucleus of approximately $10^5$ particles  clearly shows a combination of fcc, hcp and bcc crystal grains. We also  observe from \Cref{fig:spontaneous-histogram} that the fraction of bcc increases upon increasing $1/\kappa\sigma$, supporting our earlier findings from  seeding simulations.

In conclusion, we summarized our results on  crystal polymorph selection in nucleation and growth  in a  kinetic phase diagram for charged colloids. Our findings as  obtained from extensive seeding and brute-force simulations reflect the diversity of previously observed crystallization phenomena in charged colloidal suspensions. For sufficiently low supersaturations, relatively pure crystals are observed, ranging from fcc/hcp, to metastable bcc, to bcc, upon increasing  $1/\kappa\sigma$. The formation of pure metastable bcc phases are  in agreement with earlier observations from simulations and experiments \cite{xu_formation_2010, zhou_kinetics_2011, auer_crystallization_2002, kratzer_two-stage_2015, ji_crystal_2018}.

For higher  supersaturations, we find a mix of fcc, hcp, and bcc crystal grains  in a broad region of the phase diagram, in contrast to the predictions of a narrow fcc-bcc coexistence from  free-energy calculations \cite{hynninen_phase_2003}. Our findings of a mix of fcc/hcp and bcc  may explain the experimentally observed broad fcc-bcc coexistence region \cite{schope_response_1998, bareigts_packing_2020}, which may be considered as a non-equilibrium state arising by the nucleation and crystallization kinetics. Although fcc/hcp dominates here, the fraction of bcc increases with pressure and inverse screening length. 

Comparing our results with Ostwald's step rule or Alexander and McTague's conjecture demonstrates that these crystallization phenomena cannot be captured by simple rules of thumb. However, by combining our results on the nucleation barriers,  spontaneous crystallization, and seeded crystal growth, we obtain a consistent and coherent picture of the different crystallization scenarios for charged colloids. Possible extensions of our work include explicitly treating counterion screening \cite{lowen_ab_1992} or applying our methodology to oppositely charged colloids \cite{sanz_evidence_2007, sanchez-burgos_parasitic_2021}. 



\begin{acknowledgments}
 M.D. has received funding from the European Research Council (ERC) under the European Union’s
Horizon 2020 research and innovation programme (Grant
agreement No. ERC-2019-ADG 884902 SoftML).
\end{acknowledgments}

\end{document}


\title{Supplemental Material: Kinetic phase diagram for nucleation and growth of competing crystal polymorphs in charged colloidal suspensions}
\author{Willem Gispen}
\email[]{w.h.gispen2@uu.nl}

\author{Marjolein Dijkstra}
\email[]{m.dijkstra@uu.nl}

\affiliation{Soft Condensed Matter, Debye Institute for Nanomaterials Science, Utrecht University}

\maketitle

\onecolumngrid

\subsection{Code}
The code  to generate the results in this paper will be made publicly available in the Github repository \url{github.com/WillemGispen/kinetic-phase-diagram-charged-colloids}. It is straightforward to repeat the calculations for other systems of interest. We use the LAMMPS molecular dynamics code \cite{plimpton_fast_1995}.

\subsection{Molecular dynamics}
We model a charge-stabilized colloidal suspension by a system where the electrostatic interactions between the  colloids are described by  a screened-Coulomb (Yukawa) potential  
\begin{equation*}
    \beta u_Y (r) = \beta \epsilon \frac{\exp\lbrack-\kappa\sigma(r/\sigma - 1)\rbrack}{r/\sigma},
\end{equation*}
with  $\beta\epsilon$ the contact value and  $1/\kappa\sigma$  the Debye screening length. 
The excluded-volume interactions between the colloids are represented by a pseudo-hard-core potential 
\begin{equation*}
    \beta u_{PHS}(r) =\frac{2C}{3} \left( \left(\frac{\sigma}{r}\right)^{50} - \left(\frac{\sigma}{r}\right)^{49}  \right) + \frac{2}{3}, 
\end{equation*}
which reproduces well the hard-sphere equation of state using $C = 134.6$ \cite{jover_pseudo_2012}. The total  interaction potential of the pseudo-hard-core Yukawa system reads $\beta u(r) = \beta u_Y (r) + \beta u_{PHS} (r)$. Throughout this paper, we set $\beta\epsilon=81$.

We perform molecular dynamics simulations by integrating  Newton's equations of motion with the velocity-Verlet algorithm. The time step $\Delta t^*$ that we used depends on the screening length $1/\kappa\sigma$ as $\Delta t^* = 0.02/\kappa\sigma$. Here $t^*$ refers to the reduced time unit $t^* = t /\sqrt{\beta m \sigma^2}$ and $m$ is the mass of a particle. The seeding simulations are carried out in the isobaric-isothermal ensemble, i.e.\ we keep the temperature $T$ and pressure $P$ fixed  using a Nos\'e-Hoover thermostat and Nos\'e-Hoover barostat. The thermostat and barostat have a relaxation time of $100$ and $500$ time steps, respectively. The chemical potential simulations are carried out in the canonical ensemble, i.e.\ we keep the temperature $T$ and volume $V$ fixed. In addition, we use a cut-off distance of $\sigma + 12/\kappa$ for the pair potential. At this cut-off distance, the pair potential has decreased to less than $\beta u(r)=10^{-5} \beta \epsilon$. 

\subsection{Seeding}
\subsubsection{Seed preparation}
For each phase and screening length, we prepare around $16$ seeds with sizes varying between $100$ and $13 \times 10^3$ particles. We use seeds with at least $400$ particles in order to calculate the nucleation barriers, and we employ all seeds for studying crystal growth. For a target seed size $N$, we simulate a supersaturated fluid phase and a crystal phase at the same pressure $P$. Subsequently, we cut a spherical region of approximately $0.8 N$ particles from the crystal and insert it into the  fluid, such that the total system size is around twenty times as large as the target seed size. We keep the positions of the seed particles fixed and simulate the fluid for at most $8000$ MD steps or until the largest crystal cluster in the system reaches the target size $N$. We then simulate the whole system for $4000$ MD steps. If this procedure does not result in a crystal cluster within a size range of $0.9-1.1 N$  particles, we use a higher or lower pressure.


\subsubsection{Crystal nucleus size}
\label{sec:crystal-nucleus-size}
To determine the crystal nucleus size, we use the mislabeling criterion \cite{espinosa_seeding_2016} based on the locally averaged bond order parameter $\bar{q}_{6}$ \cite{steinhardt_bond-orientational_1983, lechner_accurate_2008}.

To compute $\bar{q}_{6}$, we first detect the nearest neighbors $N(i)$ of particle $i$ with the solid-angle based, nearest-neighbour algorithm  \cite{van_meel_parameter-free_2012}. Next, the local density is projected on spherical harmonics
$$
 q_{lm}(i) = \frac{1}{N(i)} \sum_{j=1}^{N(i)} Y_{lm}(\br_j - \br_i).
$$
These $q_{lm}(i)$'s are averaged over the nearest neighbors of $i$ to obtain
$$
 \bar{q}_{lm}(i) = \frac{1}{1+N(i)} \left ( q_{lm}(i) + \sum_{j=1}^{N(i)} q_{lm}(j) \right ).
$$
Finally, the averaged bond order parameter is calculated with
$$
 \bar{q}_{l}(i) = \sqrt{\frac{4\pi}{2 l + 1}\sum_{m} |\bar{q}_{lm}(i)|^2}
$$

The averaged bond order parameters $\bar{q}_{6}(i)$'s are  calculated for each particle $i$ in a bulk fluid and a bulk crystal, for a range of pressures. At each pressure, the mislabeling threshold $\bar{q}_6^*$ is determined such that the percentage of bulk solid being mislabeled as fluid is equal to the percentage of bulk fluid being mislabeled as solid, i.e.\
$$
P(\bar{q}_6(i) > \bar{q}_6^* \text{ with } i \text{ fluid}) = P ( \bar{q}_6(j) < \bar{q}_6^*  \text{ with } j \text{ solid})
$$
This mislabeling threshold $\bar{q}_6^*$ then provides a local solid-fluid classification: particles with $ \bar{q}_6 > \bar{q}_6^*$ are solid-like, and particles with $\bar{q}_6 < \bar{q}_6^*$ are fluid-like. The largest cluster of solid-like particles is found with a cut-off distance equal to the pair potential cut-off.

\subsubsection{Active learning for determining the critical pressure and critical nucleus size from seeding simulations}

To determine the critical pressure and critical nucleus size from seeding simulations, we use active learning and logistic regression. This procedure is based on the bisection method and logistic regression, taking care of stochasticity. For a fixed temperature, we first determine  the interval in which  the critical pressure $P^*$ lies. Subsequently, we repeat the following three steps:
\begin{enumerate}
    \item Estimate the critical pressure, say $P^*_{\textrm{\small est}}$, with logistic regression.
    \item Simulate the seed at the estimated pressure $P^*_{\textrm{\small est}}$ for at most $10^5$ MD steps or until it grows to $30\%$ or melts to $0.5\%$ of the system size.
    \item Add ($P^*_{\textrm{\small est}}$,$B$) to the list of observations, where $B\in\{0,1\}$ indicates whether a seed grew or melted. 
\end{enumerate}

In the first step, the estimated critical pressure  is simply the pressure for which the regression estimates a $50\%$ probability of growth. Simulating at this estimated critical pressure corresponds to the so-called `uncertainty sampling' approach to active learning \cite{settles.tr09}. Using logistic regression in this way circumvents the need to estimate the crystallization probability at a single pressure using multiple runs. The final estimate for the critical pressure is obtained after fitting $50$ observations obtained in the way just described.
 
\subsubsection{Classical nucleation theory}
We fit the obtained nucleation barriers using classical nucleation theory (CNT). From the critical nucleus size $N^*$ and the supersaturation $|\Delta\mu(P^*)|$ at the critical pressure $P^*$, we approximate the surface tension $\gamma(P^*)$ with the CNT expression
\[
 \gamma^3 = \frac{3 N^* |\Delta\mu|^3 \rho^2}{32 \pi}.
\]
Here $\rho$ is the density of the crystal phase at  critical pressure $P^*$. Next, we fit the surface tension linearly as a function of pressure, and estimate the error of this fit using Ref.\ \cite{richter_estimating_1995}.
Finally, this leads to the CNT fit
\[
\Delta G = \frac{16 \pi \gamma^3 }{3 |\Delta\mu|^2 \rho^2},
\]
which can be extrapolated to the entire metastable fluid region. The error in $\Delta G$ is estimated from the error in $\gamma$, where we neglect the errors in $\Delta \mu$ and $\rho$.


\begin{figure}[h]
  \includegraphics[width=0.6\linewidth]{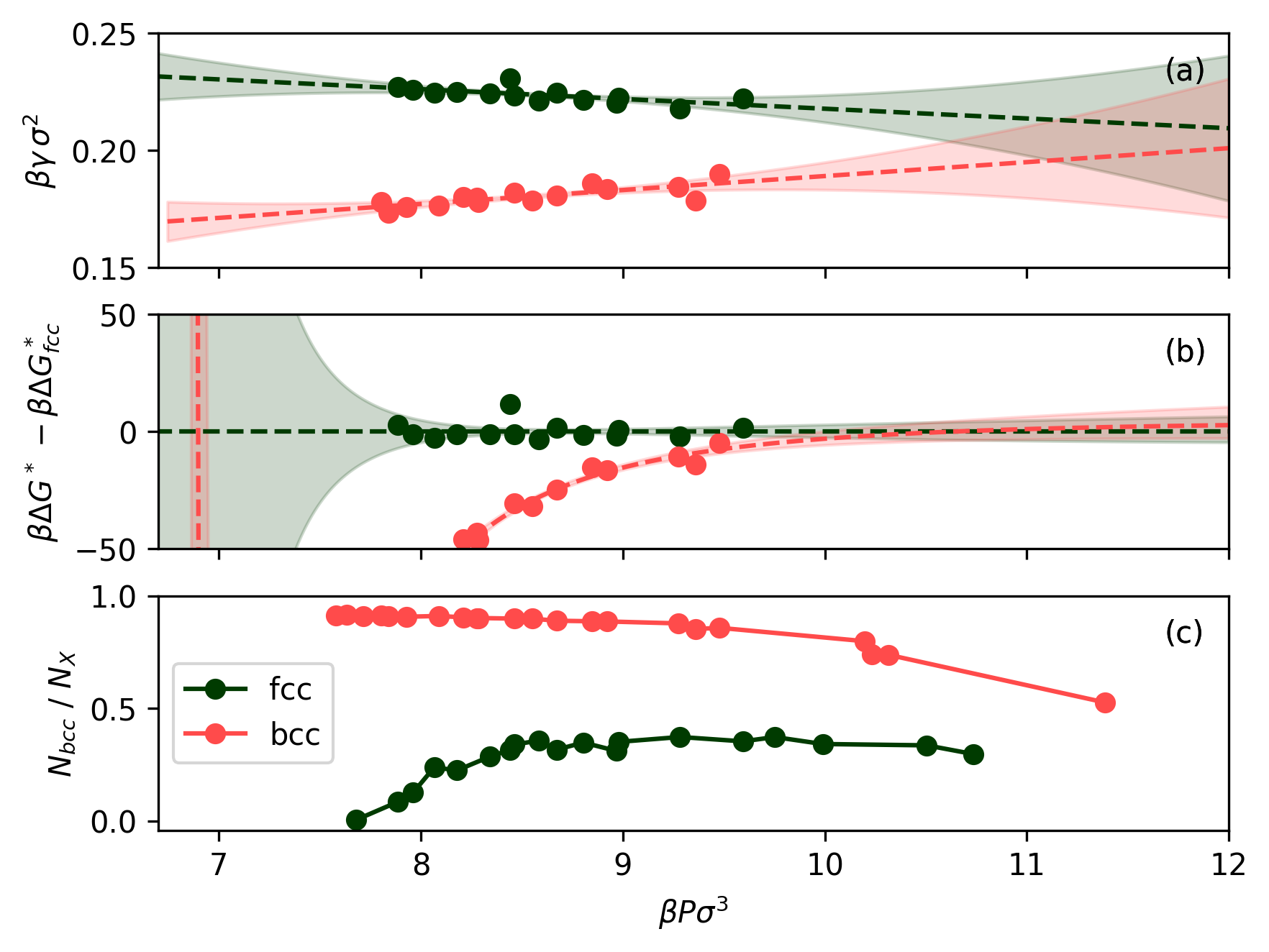}
  \caption{Results as obtained from seeding simulations of a suspension of highly charged colloids with a contact value $\beta \epsilon = 81$ and a Debye screening length $1/\kappa\sigma=0.2$. The dashed lines denote the fits using classical nucleation theory and the shaded areas denote the estimated error of the fits. (a) The solid-fluid surface tension $\gamma$ as a function of pressure $\beta P \sigma^3$, (b) difference in nucleation barrier with respect to fcc, $\Delta G^*-\Delta G^*_{fcc}$, as a function of pressure $\beta P \sigma^3$, and (c) fraction of bcc-like particles $N_{bcc}/N_x$ in the crystal growth of fcc and bcc seeds as a function of pressure $\beta P \sigma^3$ where $N_{bcc}$, $N_{fcc}$, and $N_{hcp}$  refer to the number of bcc, fcc, and hcp  particles, respectively, as recognized by polyhedral template matching, and $N_X = N_{fcc} + N_{bcc} + N_{hcp}$ is the total number of crystalline particles.  The nucleation barrier difference in (b) becomes smaller than the estimated error for $\beta P \sigma^3 > 10$. }
  \label{fig:nucleation-barrier}
\end{figure}

\subsection{Seeding simulation results for charged colloids with $1/\kappa\sigma=0.2$}
Exemplarily, we show seeding simulation results for a suspension of highly charged colloids with a contact value of $\beta\epsilon=81$ and  screening length $1/\kappa\sigma=0.2$, which is close to the triple point. In Fig. 2, we  plot the critical nucleus size $N^*$, the supersaturation $\Delta\mu$ and the nucleation barrier $\Delta G^*$. In \Cref{fig:nucleation-barrier}, we show the fluid-solid surface tension $\gamma$, the difference in nucleation barrier $\Delta G^*-\Delta G^*_{fcc}$, as well as the fraction of bcc in the crystal growth of bcc and fcc  seeds. In this region, the bcc phase is metastable with respect to  fcc, as can be seen from the lower chemical potential  for fcc with respect to bcc in Fig. 2c. However,  \Cref{fig:nucleation-barrier}a shows that bcc has a lower interfacial tension with the fluid phase than fcc. This lower surface tension causes the nucleation barrier of bcc to be lower in the pressure range {$6.9 < \beta P\sigma^3 < 10.1$}. For these pressures, the bcc seeds grow out to pure bcc crystals, whereas fcc seeds grow out into a mix of fcc, hcp and bcc (see \Cref{fig:nucleation-barrier}c of the SM).
At $\beta P\sigma^3 = 10$, the nucleation barrier difference as shown in \Cref{fig:nucleation-barrier}b becomes smaller than the estimated error. Above this pressure, the fraction of bcc in the crystal growth of bcc seeds decreases rapidly. The same trend is also observed for the other screening lengths: the pressure beyond which the nucleation barriers become indistinguishable, the `purity' of the seeds decreases rapidly.


\subsection{Free-energy calculations}
\label{sec:chemical-potential}

We compute the chemical potential of a low-density ($\rho\sigma^3=0.05$) fluid using Widom particle insertion method \cite{widom_topics_2004}. For the crystal phases, we use non-equilibrium Einstein integration \cite{freitas_nonequilibrium_2016} to determine the chemical potential at a high density (between $\rho\sigma^3=0.3$ and $\rho\sigma^3=0.6$, depending on screening length). Subsequently, we  calculate the chemical potential over the entire pressure range using thermodynamic integration of the equation of state 
%
%
%
$$
\mu (P) - \mu(P_0) = \int_{P_0}^{P} (1/\rho) \,dP.
$$
To this end, we measure the equation of state using a step size of $\Delta \rho \sigma^3 = 0.001$. For each density and phase, we simulate about $N=4000$  particles for $10^5$ simulation steps. We calculate the pressure  using only the latter half of the simulation.
%

\subsubsection{Spontaneous transformations}
During the calculations of the equation of state, we check for spontaneous phase transformations. Most importantly, the bcc and fluid phase spontaneously transform at sufficiently high  pressures. For the fluid phase, we monitor the largest crystal nucleus  size as described in Section \ref{sec:crystal-nucleus-size} of the SM. If the crystal nucleus comprises more than $5\%$ of the system size, we identify it as spontaneous nucleation. In the case of bcc, we use polyhedral template matching (PTM) \cite{larsen_robust_2016} with a root mean square deviation (RMSD) cutoff of $0.12$. If more fcc or hcp particles are recognized by PTM than bcc, we identify it as a spontantous transformation of bcc. In both cases, the lowest pressure at which the  spontaneous phase transformation occurs is used to draw the boundaries in Fig. 1.

\subsection{Kinetic phase diagram according to Ostwald's step rule}

\begin{figure}
    \centering
    \includegraphics[width=0.6\linewidth]{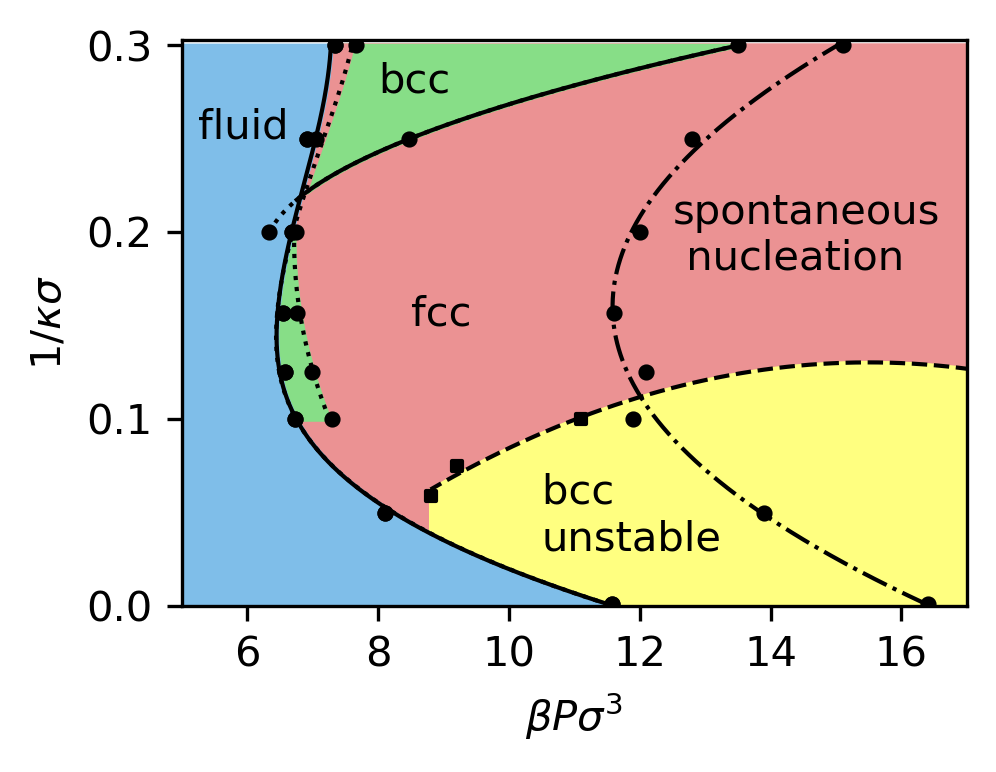}
    \caption{Kinetic phase diagram according to Ostwald's step rule. The labels fluid, bcc, fcc denote the stable fluid, bcc, fcc region. The yellow region and the dashed line denote the region where bcc is unstable as it spontaneously transforms to fcc/hcp. The dashed-dotted line denotes the pressure above which spontaneous nucleation occurs. The red and green regions denote the regions where according to Ostwald's step rule bcc and fcc, respectively, should nucleate first.}
    \label{fig:ostwald}
\end{figure}

\begin{figure}
    \centering
    \includegraphics[width=\linewidth]{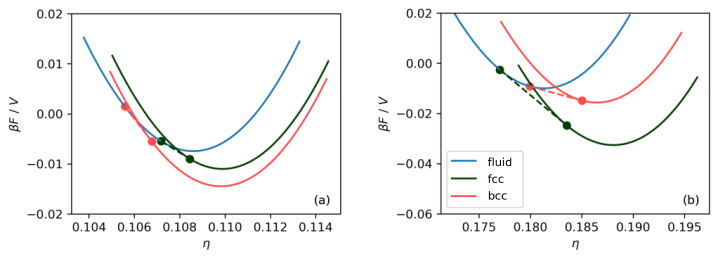}
    \caption{Helmholtz free-energy density $\beta F/V$ for a fluid,  fcc, and bcc phase of highly charged colloids with $\beta \epsilon=81$ and (a) $1/\kappa\sigma = 0.3$ and  (b)  $1/\kappa\sigma = 0.125$ as a function of packing fraction $\eta= \pi\sigma^3N/6V$. The dashed lines denote the common-tangent construction connecting the  coexisting bulk phases.}
    \label{fig:free-energy}
\end{figure}

In  \Cref{fig:ostwald}, we show the kinetic phase diagram according to Ostwald's step rule, stating that the phase that is closest in free energy to the metastable fluid phase will nucleate first. This would result in a complete reversal of the bulk phase diagram, i.e. fcc nucleates when bcc is stable, and bcc nucleates when fcc is stable except in a small region close to the binodal where fcc or bcc is even less stable than the fluid phase, as is  illustrated by the free-energy curves for the fluid, fcc, and bcc phase for $1/\kappa\sigma = 0.3$ and 0.125 in \Cref{fig:free-energy}.  

\subsection{Phase diagrams in the packing fraction - Debye screening length plane}
\begin{figure}
    \centering
    \includegraphics{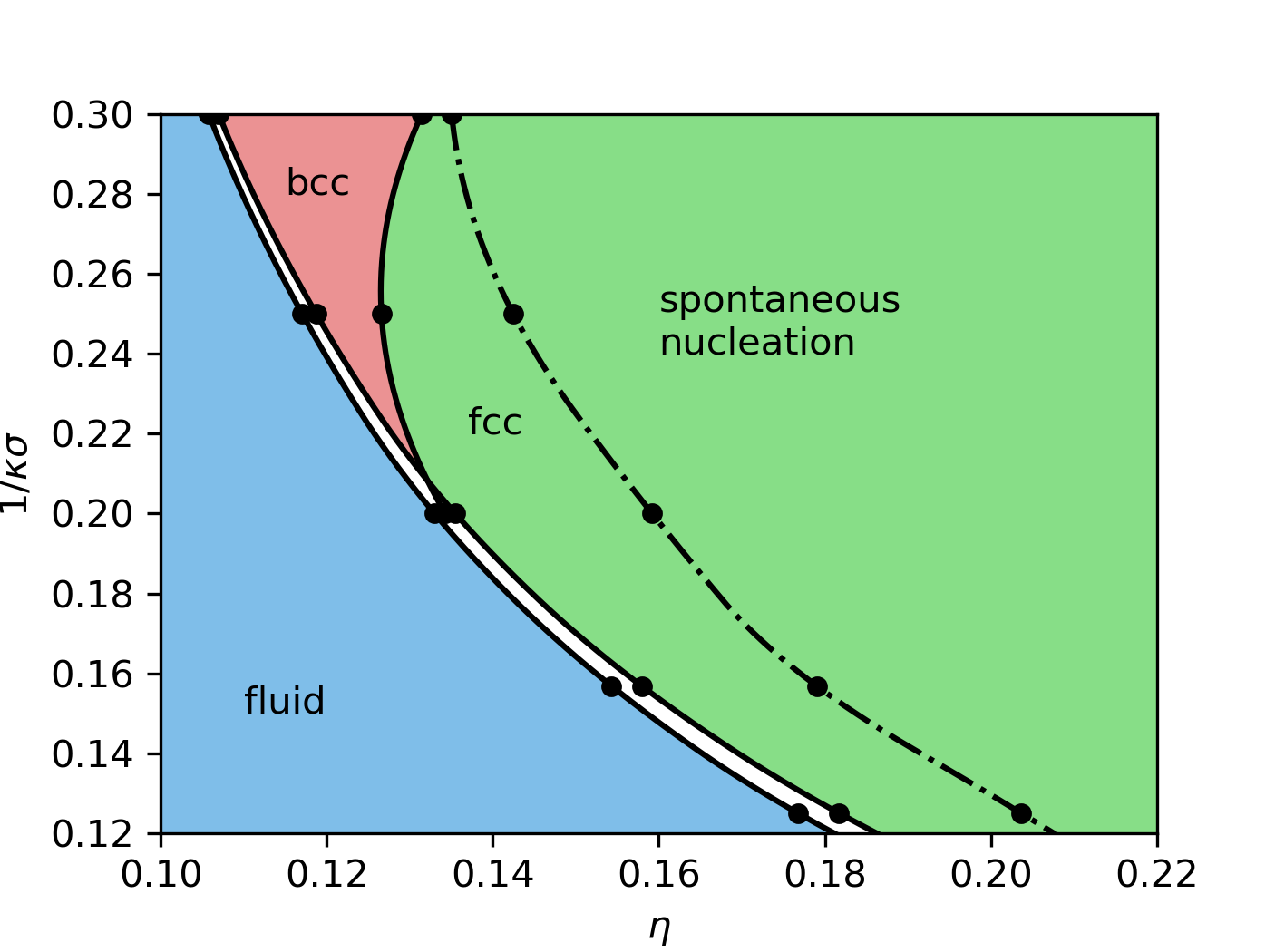}
    \caption{Bulk phase diagram of highly charged colloids with contact value $\beta\epsilon=81$ in the packing fraction $\eta$ - Debye screening length $1/\kappa\sigma$ plane. The fluid-solid and bcc-fcc binodals are denoted by solid lines, and the dash-dotted line marks the pressure beyond which the fluid spontaneously crystallizes. Dots are the actual measurements, lines are spline interpolations to guide the eye.}
    \label{fig:bulk-phase-diagram-eta}
\end{figure}

In \Cref{fig:bulk-phase-diagram-eta}, we show the bulk phase diagram in the packing fraction - Debye screening length plane, in constrast to the pressure - Debye screening length plane shown in the main text. In this presentation, we can see that spontaneous nucleation occurs for packing fractions only $2-3\%$ above the fluid-solid binodal. The fcc-bcc coexistence region is extremely narrow, in fact it is smaller than the linewidth used in the plot.

\begin{figure}
    \centering
    \includegraphics{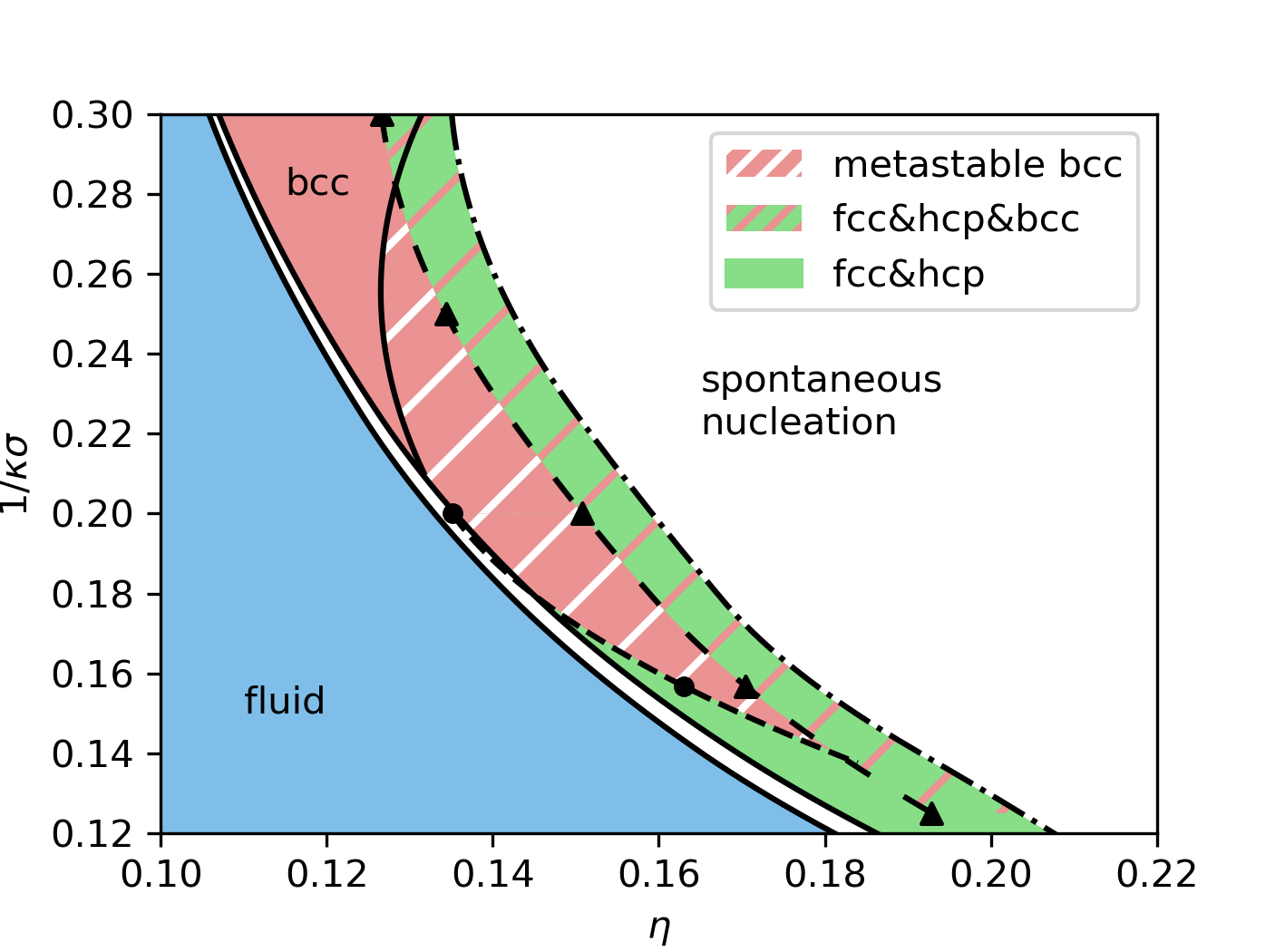}
    \caption{Kinetic phase diagram of highly charged colloids with contact value $\beta\epsilon=81$ in the packing fraction $\eta$ - Debye screening length $1/\kappa\sigma$ plane. The fluid-solid and fcc-bcc binodals are denoted by solid lines, and the dash-dotted line marks the pressure beyond which the fluid spontaneously crystallizes.  The red (green) region denotes the region where bcc (fcc) has a lower nucleation barrier than fcc (bcc). The red region with stripes near the triple point denotes the metastable bcc region, where bcc has a lower nucleation barrier while being metastable.  The triangles mark the pressures above which the nucleation barriers of fcc and bcc are indistinguishable within our statistical accuracy. Dots are actual measurements and lines are spline interpolations to guide the eye.}
    \label{fig:kinetic-phase-diagram-eta}
\end{figure}

In \Cref{fig:kinetic-phase-diagram-eta}, we also show the kinetic phase diagram in the packing fraction - Debye screening length plane. Only for packing fractions less than $2\%$ above the fluid-solid binodal, we observe (relatively) pure crystals, ranging from fcc/hcp, to metastable bcc, to bcc, upon increasing $1/\kappa\sigma$. Therefore, pure metastable bcc can only form in a narrow region near the triple point and near the fluid-solid binodal. For higher packing fractions, we find a mix of fcc, hcp and bcc crystal grains. In contrast to the extremely narrow stable fcc-bcc coexistence in \Cref{fig:bulk-phase-diagram-eta}, this non-equilibrium mix of fcc/hcp and bcc forms in a broad region of the phase diagram.

